# Quantum Sensor Duplicating the Robins Procedure

Ahmad Salmanogli

Faculty of Engineering, Electrical and Electronics Engineering Department, Çankaya University, Ankara, Turkey

Faculty of Engineering, Electrical and Electronics Engineering Department, Hacettepe University, Ankara, Turkey.

**Abstract**
In this article, we design a quantum device to duplicate the Robins procedure. The Robins use a unique method to determine the migratory direction. In the procedure utilized, the important issue is the effect of the geomagnetic field on the magnetic momentum of created radical pairs (triplet-singlet states) dancing with a special frequency. To duplicate the same operational procedure, the quantum sensor consisting of two coincident tripartite systems is designed. Each system is separately excited with the entangled photons (signal and idler) produced through nonlinearity. In a traditional tripartite system, the microwave cavity mode can be non-classically correlated with the optical cavity mode. In this study, however, there are two microwave cavities modes separately affected by the entangled photons, and these modes can be entangled. The entangled microwave photons play the same role that the triplet-singlet state of the electrons have in the Robins operating system. It is the key point that the quantum sensor is deigned to work with, in such a way that the entangled microwave photons can be strongly affected by the applied external magnetic field. In fact, it is the criterion employed by the quantum sensor to sense the magnetic field intensity and the direction. To analyze the system, the canonical conjugate method is introduced to determine the quantum sensor's Hamiltonian, and then the dynamics equations of motions are analytically derived using Heisenberg-Langevin equations.

**Introduction**
The accurate navigation ability of the European Robins has been an interesting subject for last decades [1-7]. There are some theories to clarify the Robins' procedure, however, the most important promising hypothesis to explain the phenomenon is the radical pair mechanism [1-6]. Using this phenomenon, the Robins can accurately determine its migratory direction by the effect of the geomagnetic field on the radical pairs state. In this procedure, the key point is issued from the creation of the radical pairs due to the interaction of light with a molecule at the bird's eye. These radical pairs are so important; it is because of the magnetism properties of the spin of their electrons. Quantum mechanics reveal that there are just two different states of the radical pairs. The state that two magnetic moments are opposed (↑↓) is called singlet state and when the magnetic moments are aligned (↑↑), the state is called triplet state. An interesting point happened there is that, the singlet and triplet states are converted to each other with a special frequency, and that is so-called state dancing [1, 2]. It is contributed to the fact that two states have same energy, and more technically, it is a consequence of the coherence of the electron spins. Several studies have reported that the spin coherence time is on order of microsecond, and it is reciprocally equivalent to the electron Larmor frequency 1.4 MHz [4]. The migratory direction detection is made when the geomagnetic field disturbs the singlet-triplet dancing; for example, the timing of the state dancing is manipulated. Additionally, there are some other studies that have been investigated to clarify the different feature of the Robins; for instance, in [1] the radio frequency effect is studied on the Robins migratory detection procedure. It is shown that if the frequency of the applied external magnetic field matches with the dancing frequency of the electron pairs, the operation of the Robins is distorted [1]. Moreover, it is shown in [5] that the red light can disturb the migratory bird's orientation. The other work investigates the quantum coherence role and entanglement in the chemical compass [7]. In this work, it is shown that the singlet and triplet states are entangled, and the geomagnetic field can disturb the non-local and non-trivial behavior of the states. Also, with the knowledge of the Robins interesting procedure, some researches have focused on establishing the device to detect the external weak geomagnetic field [8, 9]. Utilizing the above suggested points, the present study aims to design a quantum device to duplicate the Robins procedure as exactly as possible. The device utilizes the entanglement phenomenon to accurately detect the external magnetic field effect. Similar to the Robins procedure, in this device, the entanglement between microwave cavities modes are affected by the external magnetic field.

The quantum sensor, consisting of two independent tripartite systems (so-called S_I and S_II), which are able to generate two entangled photons in the microwave frequency. In general, a tripartite system contains optical cavity (OC), microresonator (MR), and microwave cavity (MC), in which the cavity modes are coupled to each other. In a traditional tripartite system, it has been shown that there is a possibility to generate entanglement states between the OC and MC modes [10-13]. In contrast to those systems, in this study, the entangled photons [14-16] are initially employed to separately excite the coincident tripartite systems. Due to the excitation through the entangled photons, it is shown that at some special detuning frequencies, the entangled microwave modes ($MC_1$-$MC_2$) are achieved. The entangled microwave photons play the same role that the triplet-singlet state of electrons plays in the Robins procedure. The microwave entangled photons are generated through two physically un-coupled LC circuits which are separately stimulated by entangled photons. Additionally, it is shown that the sensor has the ability to sense the external magnetic field changing through which the entanglement between microwave photons is manipulated. It is like the case that the Robins use to determine its migratory route. Most importantly, compared to the Robins, the designed sensor is not affected by other radio frequency (RF) sources operating in the range of $f_{ext}$ <= 1.6 GHz, where $f_{ext}$ is the RF source frequency. The applied frequency higher than 1.6 GHz disturbs the quantum sensor operation, whereas the Robins operation is disturbed in the range of Larmor frequency [1]. The system description section will introduce the details of how the sensor is able to determine the external magnetic field intensity and its direction. It should be noted that, in this study, the external magnetic field is sensed by a Hall sensor [17, 18]. Ultra-high sensitive Hall sensors show sensitivity around 10 mT, which is approximately 200 times greater than the typical geomagnetic field intensity. Therefore, geomagnetic field (22 μT to 67 μT) cannot be easily sensed by a traditional Hall sensor. Nonetheless, the main aim of this study is not to sense the geomagnetic weak field. We just focus to design a quantum sensor utilizing the entanglement between microwave photons which is manipulated due to change of the intensity and direction of the external magnetic field. Certainly, one can use the same techniques utilized in [8, 9] to sense the geomagnetic field and improve the sensor performance. Finally, for system analysis, the quantum sensor total Hamiltonian is derived, and then the dynamics equations of motions are analytically examined using Heisenberg-Langevin equations. Notably, for each of the tripartite system, the contributed Hamiltonian is derived from the system's total Lagrangian.

**Theory and Background**

*A. System description*

Scheme 1 schematically illustrates the quantum sensor. It contains two coincident tripartite systems, including OC, MR and MC. Each tripartite system is separately excited by entangled photons, one of them is excited with the signal and the other is excited with the idler. The signal and idler are initially generated through interaction of a high intensity laser with OPDC [14-16]. It has been examined that two generated photons are entangled $\omega_p = \omega_{os} + \omega_{oi}$, where $\omega_p$, $\omega_{os}$, $\omega_{oi}$ are the pump, signal and idler angular frequencies, respectively. After excitation of optical cavities, the contributed modes are coupled to MR through the optical pressure, then MR oscillator is directly coupled to MC by changing the capacitance in the circuit. Generally, the tripartite system is operated based on the cavities modes coupling to each other [10-13]. However, in this system, it needs few additional electrical circuits to sense the external magnetic field and transfer its effect to the related tripartite system. In other words, the tripartite cavities modes coupling features are manipulated by applying an external magnetic field. In Scheme 1, a magnet is used to illustrate the external magnetic field direction in which S-pole influences the S_II. To sense the magnetic field intensity, traditionally, an ultra-high sensitive Hall sensor [17, 18] is utilized. These sensors show sensitivity around 10 mT. Therefore, the designed sensor works on the area that Hall sensors are able to sense the magnetic field. Another electronic element used in this system is the Varactor diode. It senses the Hall sensor's voltage, and the contributed capacitance $C_{VS}$ is changed based on the voltage drop across. In fact, the magnetic field amplitude is expressed in terms of the Varactor diode's capacitance alteration through the simple electronic circuit. Notably, each tripartite system contains two capacitances, one of them $C_s(x)$ or $C_i(x)$ is changed regarding the optical pressure applied on the MR oscillator, and the other $C_{VS}$ is used to detect the external magnetic field. Clearly, at each point in space, the magnetic field direction is unique, therefore, at each point in space, only one of the Hall sensors in the system is affected by the S-pole of the magnetic field. In other words, only one of the Varactor diodes at each side provides a considerable capacitance regarding the voltage drop across it. By changing the one side capacitance, the contributed LC circuit electrical properties are strongly changed. It is the point utilizing to manipulate the entanglement between $MC_1$ and $MC_2$, where are the microwave cavity modes of S_I and S_II, respectively.

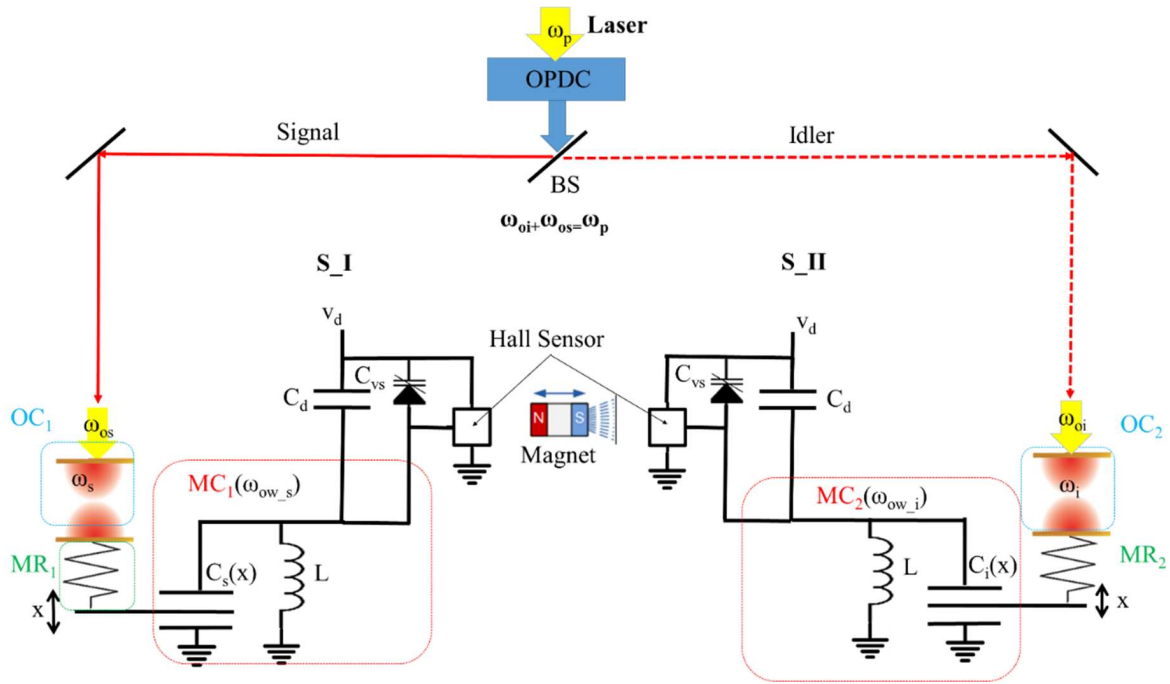

Scheme. 1 Quantum sensor consisting of two coincident tripartite systems (pair) to be excited with signal (left) and idler (right), and additional electrical circuits with Hall sensor to detect the external magnetic field intensity; on a magnetic line in space at each time just one of the Hall sensors can sense the S-pole effect.

The sensor illustrated in Scheme 1 is used only to sense the intensity of the applied magnetic field. In other words, it cannot determine the magnetic field direction. To detect the magnetic field direction, we design an array of the pairs illustrated in Scheme 2a. It works like a sensitive compass. The direction detection of the magnetic field can be possible using this arrangement. For instance, if the magnetic field is in the direction of the illustrated compass in the figure, then the red pair is activated, and the red pair microwave photons ($MC_1$ and $MC_2$) becomes entangled. Therefore, based on the design, when one of the Hall sensors senses an intensity above the threshold, the microwave photons of the activated pair's become entangled. Nonetheless, the MC modes of other pairs in the array cannot be entangled. The blue-colored pair, for example, can sense the attenuated magnetic field intensity based on the direction of the illustrated compass, which means that the MC modes of this pair is separable. Another interesting point about the system is that two different pairs cannot be coupled to each other. For instance, S_I from the red pair cannot be entangled with S_II from the blue pair. This is because the excitation source of the different pairs is different; the signal source to excite the red pair is not entangled with the signal source of the blue one. Additionally, this figure depicts two different angles $\alpha_s$ and $\alpha_e$, which are the subjected area and the effective area of the external magnetic field, respectively. It is clearly seen from figure that the magnetic field can influence more than one pair with the angle of $\alpha_s$, while the results show that the $MC_1$ and $MC_2$ entanglement occurs for one pair just covered by $\alpha_e$. This is because the amplitude of the magnetic field is maximized only in the effective area $\alpha_e$, not in $\alpha_s$. Furthermore, Scheme 2b depicts the schematic of the detection of the magnetic field direction by the designed sensor. It is displayed that the intensity of the magnetic field is maximized at the direction of arrows. In other words, on each line, the maximum amount of the magnetic field intensity (S-pole) is absolutely sensed by one pair.

An important question is that, is there a chance to sense a common magnetic field by two different pairs? One can consider very close pairs in the array to accurately determine the magnetic field direction. In this case, it is possible to approximately sense the same field intensity by too close adjacent pairs. It does not disturb the system performance, since the indicated pairs are very close to each other and indicate approximately one direction. More importantly, it is clear that the microwave cavity mode of different pairs is not able to nonclassicaly couple to each other. It is the main point causing this sensor to operate uniquely and safely to determine the direction of the magnetic field.

Table. 1 data used to simulate the quantum sensor

| | |
|---|---|
| $\alpha_c$ | 0.012 |
| $\lambda_p$ | $405*10^{-9}$ m |
| $\gamma$ | 500 Hz |
| m | $90*10^{-9}$ g |
| L | $5*10^{-12}$ H |
| $\kappa_s = \kappa_i$ | $0.05\omega_m$ |
| $\kappa_{cs} = \kappa_{ci}$ | $0.02\omega_m$ |
| $\omega_m$ | $2\pi*10^6$ Hz |
| $C_1(x_0)$ | $60*10^{-12}$ F |
| $C_d$ | $0.5*10^{-12}$ F |
| $C_{vs}$ | $(10\sim120)*10^{-12}$ F |

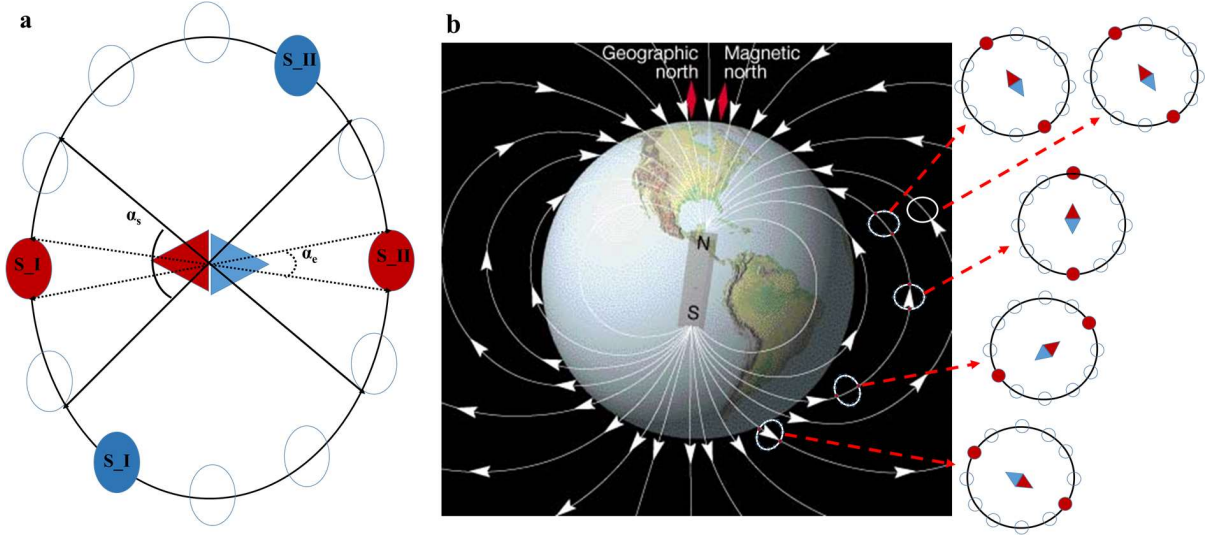

Scheme 2. a) Schematic of quantum sensor to detect the magnetic field direction at which no pairs are affected by others. Based on the illustrated compass, the red pair is only one that is excited by the magnetic field; also a brief comparing is made between $\alpha_s$ and $\alpha_e$; $\alpha_s$ is the angle to cover the external magnetic field, while $\alpha_e$ is the angle to indicate the effective area at which the sensor is designed to safely operate in; b) schematic of detection of the magnetic field direction by the designed quantum sensor.

### B. Quantum sensor dynamics of motions (one pair)

After a short discussion about the general operation of the system, in this section, the system is theoretically analyzed. The canonical conjugate method [19] in which the matter is considered by the related harmonic polarization field is applied. It is mentioned that the system illustrated in scheme. 1 contains two similar tripartite systems; then in the following, the total Hamiltonian is calculated for one of the tripartite system. Following the standard approach in quantum electrodynamics, we start with the system's total Lagrangian represented in terms of the conjugate variables as:

$$L_{total} = L_{OC} + L_{MR} + L_{MC} + L_{OC-MR}$$

$$L_{OC} = \frac{\varepsilon_0}{2}(E^2 - A^2)$$

$$L_{MR} = \frac{P^2}{2m} + \frac{\alpha_c PA}{m} + \frac{\alpha_c^2 A^2}{m} - \frac{m\omega_m^2 X^2}{2} \quad (1)$$

$$L_{MC} = \frac{Q^2}{2C_N} - \frac{\phi^2}{2L} + \frac{C_1(x)C_M V_d^2}{2C_N}$$

$$L_{OC-MR} = \frac{-\alpha_c PA}{m} - \frac{\alpha_c^2 A^2}{m}$$

where $L_{OC}$, $L_{MR}$, $L_{MC}$, and $L_{OC-MR}$ are optical cavity, microresonator, microwave cavity, and optical cavity-microresonator interaction Lagrangian, respectively. Also, $\varepsilon_0$, **E**, **A**, **P**, **X**, $\alpha_c$, $\omega_m$, **L**, **Q**, **Φ**, $V_d$ and m are the free space dielectric constant, electric field operator, vector potential operator, MR oscillator momentum operator, MR oscillator position operator, OC and MR interaction coefficient factor, MR oscillation angular frequency, LC circuit inductance, the stored charge operator in the MC capacitor, magnetic flux operator, MC input driving field, and MR resonator mass, respectively. Moreover, $C_M = C_d + C_{VS}$ and $C_N = C_M + C_1(x)$ where $C_1(x_0) = \varepsilon A_{cap}/d_{cap}$. Also, $A_{cap}$ and $d_{cap}$ are the capacitor's plates surface area and the separation between, respectively. It should be noted that the third term in $L_{MC}$ denotes the interaction Lagrangian between MC and MR. Using Eq. 1, the Hamiltonians of the system are expressed as:

$$H_{OC} = \frac{\varepsilon_0}{2}(\hat{E}^2 + \omega_{os}^2 \hat{A}^2)$$

$$H_{MR} = \frac{1}{2m}\hat{p}^2 + \frac{m\omega_m^2}{2}\hat{x}^2$$

$$H_{MC} = \frac{1}{2C_N}\hat{Q}^2 + \frac{1}{2L}\hat{\phi}^2 + \frac{C_M V_d}{2C_{N0}}\hat{Q} \qquad (2)$$

$$H_{OC-MR} = \frac{\alpha_c}{m}\hat{p}\hat{A}$$

$$H_{MC-MR} = \frac{-C_M}{2C_{N0}^2}\frac{dC_1(x)}{dx}\frac{1}{2C_M}\hat{Q}^2 \hat{x}$$

Where $\omega_{os}$, and $H_{MC-MR}$ are the optical cavity resonance frequency, and interaction Hamiltonian between MC and MR, respectively. Moreover, $C_{N0} = C_M + C_1(x_0)$ where $x_0$ indicates the equilibrium position of MR oscillator. After calculation of the cavities creation and annihilation operators and substituting in Eq. 2, the simplified system total Hamiltonian is given by:

$$H_{OC\_s1} = \hbar\omega_{os}\hat{a}_s^+ \hat{a}_s \qquad\qquad H_{OC\_s2} = \hbar\omega_{oi}\hat{a}_i^+ \hat{a}_i$$

$$H_{MR\_s1} = \frac{\hbar\omega_{ms}}{2}(\hat{p}_{0s}^2 + \hat{x}_{0s}^2) \qquad H_{MR\_s2} = \frac{\hbar\omega_{mi}}{2}(\hat{p}_{0i}^2 + \hat{x}_{0i}^2)$$

$$H_{MC\_s1} = \hbar\omega_{\omega s}\hat{c}_s^+\hat{c}_s - jV_d C_M\sqrt{\frac{\hbar\omega_{\omega s}}{2C_{N0}}}(\hat{c}_s - \hat{c}_s^+) \quad H_{MC\_s2} = \hbar\omega_{\omega i}\hat{c}_i^+\hat{c}_i - jV_d C_M\sqrt{\frac{\hbar\omega_{\omega i}}{2C_{N0}}}(\hat{c}_i - \hat{c}_i^+)$$

$$H_{OC-MR\_s1} = \hbar\sqrt{\frac{\alpha_c^2 \omega_{ms}}{2\varepsilon_0 \omega_{os} m}}(\hat{a}_s^+ + \hat{a}_s)\hat{p}_{0s} \qquad H_{OC-MR\_s2} = \hbar\sqrt{\frac{\alpha_c^2 \omega_{mi}}{2\varepsilon_0 \omega_{oi} m}}(\hat{a}_i^+ + \hat{a}_i)\hat{p}_{0i} \qquad (3)$$

$$H_{MC-MR\_s1} = \frac{-\hbar\omega_{\omega s}}{2C_{N0}}\frac{dC_1(x)}{dx}\sqrt{\frac{\hbar}{\omega_{ms} m}}\hat{x}_{0s}\hat{c}_s^+\hat{c}_s \quad H_{MC-MR\_s2} = \frac{-\hbar\omega_{\omega i}}{2C_{N0}}\frac{dC_1(x)}{dx}\sqrt{\frac{\hbar}{\omega_{mi} m}}\hat{x}_{0i}\hat{c}_i^+\hat{c}_i$$

$$H_{oc-drive\_s1} = i\hbar E_c(\hat{a}_s^+ e^{(-j0.5\omega_p t)} - \hat{a}_s e^{(j0.5\omega_p t)}) \quad H_{oc-drive\_s2} = i\hbar E_c(\hat{a}_i^+ e^{(-j0.5\omega_p t)} - \hat{a}_i e^{(j0.5\omega_p t)})$$

$$H_{int-eph} = -\hbar\frac{\chi^{(2)}\sqrt{\omega_{\omega s}\omega_{\omega i}}}{2n^2}E_{p0}[\hat{a}_s^+\hat{a}_i^+ + \hat{a}_s\hat{a}_i]$$

where $(a_j^+, a_j)$, and $(c_j^+, c_j)$, with subscript j = s or i, are the creation and annihilation operators for the optical and microwave cavities, respectively. Also, $\omega_{\omega j}$, $p_{0j}$, $x_{0j}$, $E_c$, $E_{p0}$, n, $\chi^{(2)}$, and h are the MC cavity resonance frequency, normalized quadrature momentum and position operators, OC cavity input driving rate, input laser amplitude to excite the OPDC, nonlinear material refractive index, nonlinear material susceptibility, Plank's constants, respectively. Notably, $H_{int-sph}$ is the only part of the system Hamiltonian that connect the two tripartite systems, S_I and S_II, each other through the non-classical connection between the entangled photons (signal and idler). It is suggesting that if $H_{int-sph}$ goes to be vanished, for example by decreasing of $\chi^{(2)}$, the tripartite systems modes either optical modes or microwave modes remained separable. This point will be discussed in the following section.

Using the Hamiltonian defined in Eq. 3, and by applying Heisenberg-Langevin equations [24, 25], the dynamics equations of motions are derived for one pair of the quantum sensor. It has to consider the effect of the damping rate and noise which is due to the interaction of the sensor with the environment. The related dynamics of equation of motions are presented as:

$$\dot{a}_s = -(i\Delta_{ocs} + \kappa_s)a_s - iG_{11}p_s - iG_{int}E_{p0}a_i^+ + E_c + \sqrt{2\kappa_s}\,a_{in}$$

$$\dot{a}_i = -(i\Delta_{oci} + \kappa_i)a_i - iG_{22}p_i - iG_{int}E_{p0}a_s^+ + E_c + \sqrt{2\kappa_i}\,a_{in}$$

$$\dot{c}_s = -(i\Delta_{o\omega s} + \kappa_{cs})c_s + i0.5\Delta_{o\omega 1}G_{13}x_s c_s + E_\omega + \sqrt{2\kappa_{cs}}\,c_{in}$$

$$\dot{c}_i = -(i\Delta_{o\omega i} + \kappa_{ci})c_i + i0.5\Delta_{o\omega 2}G_{23}x_i c_i + E_\omega + \sqrt{2\kappa_{ci}}\,c_{in} \quad (4)$$

$$\dot{x}_s = \omega_{ms}p_s + G_{11}(a_s^+ + a_s)$$

$$\dot{x}_i = \omega_{mi}p_i + G_{22}(a_i^+ + a_i)$$

$$\dot{p}_s = -\gamma_s p_s - \omega_{ms}x_s + 0.5\Delta_{o\omega s}G_{13}c_s^+ c_s + b_{in}$$

$$\dot{p}_i = -\gamma_i p_i - \omega_{mi}x_i + 0.5\Delta_{o\omega i}G_{23}c_i^+ c_i + b_{in}$$

where $\kappa_j$, $\gamma_j$, and $\kappa_{cj}$, with subscript j = s or i are the optical cavity, microresonator, microwave cavity damping rate, respectively. Also, $\Delta_{ocj}$, and $\Delta_{o\omega j}$ are the related detuning frequency, and $G_{11} = \sqrt{2}(\alpha_c^2\omega_{ms}/2\varepsilon_0 m\omega_{os})$, $G_{22} = \sqrt{2}(\alpha_c^2\omega_{mi}/2\varepsilon_0 m\omega_{oi})$, $G_{13} = (C_{1s}'(x)/C_{N0})\sqrt{(\hbar/m\omega_{ms})}$, $G_{23} = (C_{1i}'(x)/C_{N0})\sqrt{(\hbar/m\omega_{mi})}$, and $G_{int} = -\chi^{(2)}E_{p0}\sqrt{(\omega_{os}\omega_{oi})}/2n^2$. In these relations, $C_{1s}'(x)$ and $C_{1i}'(x)$ denote the change of the MC circuit capacitance with respect to the signal and idler optical pressure, respectively.

A simple way to achieve stationary and robust entanglement in continuous variable, is to determine an operating point at which the cavities are driven. Under the condition that the interaction field is so strong, one can focus on the linearization and calculate the quantum fluctuation around the semi-classical fixed point [10-13]. Therefore, the cavity modes can be written as the superposition of a fix point and the contributed fluctuations as: **a$_s$** = A$_s$+**δa$_s$**, **a$_i$** = A$_i$+**δa$_i$**, **c$_s$** = C$_s$+**δc$_s$**, **c$_i$** = C$_i$+**δc$_i$**, **x$_s$** = q$_s$+**δx$_s$**, **x$_i$** = q$_i$+**δx$_i$**, **p$_s$** = P$_s$+**δp$_s$**, and **p$_i$** = P$_i$+**δp$_i$**, where the capital alphabet denotes the stationary point, and δ indicates the fluctuation around the fix point. In the steady state condition, Eq. 4 becomes:

$$0 = -(i\Delta_{ocs} + \kappa_s)A_s - iG_{11}P_s - iG_{int}E_{p0}A_i^* + E_c$$

$$0 = -(i\Delta_{oci} + \kappa_i)A_i - iG_{22}P_i - iG_{int}E_{p0}A_s^* + E_c$$

$$0 = -(i\Delta_{o\omega s} + \kappa_{cs})C_s + i0.5\Delta_{o\omega 1}G_{13}q_s C_s + E_\omega$$

$$0 = -(i\Delta_{o\omega i} + \kappa_{ci})C_i + i0.5\Delta_{o\omega 2}G_{23}q_i C_i + E_\omega \quad (5)$$

$$0 = \omega_{ms}p_s + G_{11}(A_s^* + A_s)$$

$$0 = \omega_{mi}p_i + G_{22}(A_i^* + A_i)$$

$$0 = -\gamma_s p_s - \omega_{ms}q_s + 0.5\Delta_{o\omega s}G_{13}|c_s|^2$$

$$0 = -\gamma_i p_i - \omega_{mi}q_i + 0.5\Delta_{o\omega i}G_{23}|c_i|^2$$

To solving Eq. 5, it is assumed that $A_s$ and $A_i$ is a real number, and $Re\{A_s\}\gg1$, $Re\{A_i\}\gg1$ and also $|C_s|\gg1$ and $|C_i|\gg 1$. It should be noted that the fluctuation is calculated around the fix point, and the selection of such those fix points don't affect the mode's fluctuation. Finally, the dynamics equations of the modes fluctuations are given by:

$$\dot{\delta a_s} = -(i\Delta_{ocs}+\kappa_s)\delta a_s - iG_{11}\delta p_s - iG_{int}E_{p0}\delta a_i^+ + E_c + \sqrt{2\kappa_s}\delta a_{in}$$

$$\dot{\delta a_i} = -(i\Delta_{oci}+\kappa_i)\delta a_i - iG_{22}\delta p_i - iG_{int}E_{p0}\delta a_s^+ + E_c + \sqrt{2\kappa_i}\delta a_{in}$$

$$\dot{\delta c_s} = -(i\Delta_{o\omega s}+\kappa_{cs})\delta c_s + i0.5\Delta_{o\omega 1}G_{13}\{q_s\delta c_s + \delta x_s C_s\} + E_\omega + \sqrt{2\kappa_{cs}}\delta c_{in}$$

$$\dot{\delta c_i} = -(i\Delta_{o\omega i}+\kappa_{ci})\delta c_i + i0.5\Delta_{o\omega 2}G_{23}\{q_i\delta c_i + \delta x_i C_i\} + E_\omega + \sqrt{2\kappa_{ci}}\delta c_{in} \quad (6)$$

$$\dot{\delta x_s} = \omega_{ms}\delta p_s + G_{11}(\delta a_s^+ + \delta a_s)$$

$$\dot{\delta x_i} = \omega_{mi}\delta p_i + G_{22}(\delta a_i^+ + \delta a_i)$$

$$\dot{\delta p_s} = -\gamma_s\delta p_s - \omega_{ms}\delta x_s + 0.5\Delta_{o\omega s}G_{13}\{C_s^*\delta c_s + \delta c_s^+ C_s\} + \delta b_{in}$$

$$\dot{\delta p_i} = -\gamma_i\delta p_i - \omega_{mi}\delta x_i + 0.5\Delta_{o\omega i}G_{23}\{C_i^*\delta c_i + \delta c_i^+ C_i\} + \delta b_{in}$$

To study the entanglement between cavities modes, Eq. 6 should be solved. For simplicity, the matrix form of the appropriate quadrature operators of the intra-cavity fields are introduced as:

$$\begin{bmatrix} \dot{\delta x_s} \\ \dot{\delta p_s} \\ \dot{\delta x_i} \\ \dot{\delta p_i} \\ \dot{\delta x_{os}} \\ \dot{\delta y_{os}} \\ \dot{\delta x_{oi}} \\ \dot{\delta y_{oi}} \\ \dot{\delta x_{\omega s}} \\ \dot{\delta y_{\omega s}} \\ \dot{\delta x_{\omega i}} \\ \dot{\delta y_{\omega i}} \end{bmatrix} = \underbrace{\begin{bmatrix} 0 & \omega_{ms} & 0 & 0 & G_{11} & 0 & 0 & 0 & 0 & 0 & 0 & 0 \\ -\omega_{ms} & -\gamma & 0 & 0 & 0 & 0 & 0 & 0 & M_1 & M_2 & 0 & 0 \\ 0 & 0 & 0 & \omega_{mi} & 0 & 0 & G_{22} & 0 & 0 & 0 & 0 & 0 \\ 0 & 0 & -\omega_{mi} & -\gamma & 0 & 0 & 0 & 0 & 0 & 0 & M_3 & M_4 \\ 0 & 0 & 0 & 0 & -\kappa_s & \Delta_{ocs} & 0 & -G_{int} & 0 & 0 & 0 & 0 \\ 0 & -G_{11} & 0 & 0 & -\Delta_{ocs} & -\kappa_s & -G_{int} & 0 & 0 & 0 & 0 & 0 \\ 0 & 0 & 0 & 0 & 0 & -G_{int} & -\kappa_i & \Delta_{oci} & 0 & 0 & 0 & 0 \\ 0 & 0 & 0 & -G_{22} & -G_{int} & 0 & -\Delta_{oci} & -\kappa_i & 0 & 0 & 0 & 0 \\ -M_2 & 0 & 0 & 0 & 0 & 0 & 0 & 0 & \substack{M_6 \\ -\kappa_{cs}} & \substack{M_5 \\ +\Delta_{o\omega s}} & 0 & 0 \\ M_1 & 0 & 0 & 0 & 0 & 0 & 0 & 0 & \substack{-M_5 \\ -\Delta_{o\omega s}} & \substack{M_6 \\ -\kappa_{cs}} & 0 & 0 \\ 0 & 0 & -M_4 & 0 & 0 & 0 & 0 & 0 & 0 & 0 & \substack{M_8 \\ -\kappa_{ci}} & \substack{M_7 \\ +\Delta_{o\omega i}} \\ 0 & 0 & M_3 & 0 & 0 & 0 & 0 & 0 & 0 & 0 & \substack{-M_7 \\ -\Delta_{o\omega i}} & \substack{M_8 \\ -\kappa_{ci}} \end{bmatrix}}_{A_{i,j}} \times \underbrace{\begin{bmatrix} \delta x_s \\ \delta p_s \\ \delta x_i \\ \delta p_i \\ \delta x_{os} \\ \delta Y_{os} \\ \delta x_{oi} \\ \delta Y_{oi} \\ \delta x_{\omega s} \\ \delta Y_{\omega s} \\ \delta x_{\omega i} \\ \delta Y_{\omega i} \end{bmatrix}}_{u(0)} + \underbrace{\begin{bmatrix} 0 \\ \delta b_n \\ 0 \\ \delta b_n \\ \sqrt{2\kappa_s}\delta x^c \\ \sqrt{2\kappa_i}\delta y^c \\ \sqrt{2\kappa_s}\delta x^c \\ \sqrt{2\kappa_i}\delta y^c \\ \sqrt{2\kappa_{cs}}\delta x^\omega \\ \sqrt{2\kappa_{ci}}\delta y^\omega \\ \sqrt{2\kappa_{cs}}\delta x^\omega \\ \sqrt{2\kappa_{ci}}\delta y^\omega \end{bmatrix}}_{n(t)} \quad (7)$$

where $M_1 = 0.5\sqrt{2}*G_{13}\Delta_{o\omega s}Re\{C_s\}$, $M_2 = 0.5\sqrt{2}*G_{13}\Delta_{o\omega s}Im\{C_s\}$, $M_3 = 0.5\sqrt{2}*G_{23}\Delta_{o\omega i}Re\{C_i\}$, $M_4 = 0.5\sqrt{2}*G_{23}\Delta_{o\omega i}Im\{C_i\}$, $M_5 = -0.5*G_{13}\Delta_{o\omega s}Re\{q_s\}$, $M_6 = -0.5*G_{13}\Delta_{o\omega s}Im\{q_s\}$, $M_7 = -0.5*G_{23}\Delta_{o\omega i}Re\{q_i\}$, $M_8 = -0.5*G_{23}\Delta_{o\omega i}Im\{q_i\}$. Also, in Eq. 7 $\delta x^c$, $\delta y^c$, $\delta x^\omega$, and $\delta y^\omega$ are the quadrature operator of the related noises. Eq. 7 can be simply solved and yield to a general form as $u(t) = \exp(A_{i,j}t)u(0) + \int(\exp(A_{i,j}s).n(t-s))ds$, where n(s) is the noise column matrix. The related input noises obey the following correlation function [10, 12, 22].

$$<a_{in}(s)a_{in}^*(s')>=[N(\omega_c)+1]\delta(s-s'); <a_{in}^*(s)a_{in}(s')>=[N(\omega_c)]\delta(s-s')$$
$$<c_{in}(s)c_{in}^*(s')>=[N(\omega_\omega)+1]\delta(s-s'); <c_{in}^*(s)c_{in}(s')>=[N(\omega_\omega)]\delta(s-s') \quad (8)$$
$$<b_{in}(s)b_{in}^*(s')>=[N(\omega_m)+1]\delta(s-s'); <b_{in}^*(s)b_{in}(s')>=[N(\omega_m)]\delta(s-s')$$

where $N(\omega) = [\exp(\hbar\omega/k_B T)-1]^{-1}$, $k_B$ and T are the Boltzmann's constant and operational temperature, respectively [10, 12]. Indeed, $N(\omega)$ is the equilibrium mean thermal photon numbers of the different modes. After solving Eq. 7, it gives the different cavity modes fluctuation operator, then one can calculate the entanglement between considered modes operators. It is mentioned that in such this sensor, the entanglement between two microcavity modes are essential. Therefore, in this article, we just mathematically focus on the $MC_1$-$MC_2$ entanglement. To study the entanglement between two modes, it is utilized Simon-Peres-Horodecki criterion for continuous state separability which is concluded from the uncertainty principle [20-23] as:

$$\lambda_{SPH} = \det(A).\det(B)+(0.25-|\det(C)|)^2 - tr(AJCJBJC^T J)-0.25\times(\det(A)+\det(B))\geq 0 \quad (9)$$

where J = [0, 1; -1, 0] and notation "tr" stands for the trace of the matrix. This criterion is a necessary and sufficient condition of bipartite Gaussian states system separability. So, if it is assumed that the states in the designed system are Gaussian, it can normally use this criterion, Eq. 9, to study the entanglement between modes. Also, A, B, and C are 2×2 correlation matrix elements [A, C; $C^T$, B] that, can be represented for the case of $MC_1$-$MC_2$ as:

$$A = \begin{bmatrix} <\delta x_{\omega s}^2>-<\delta x_{\omega s}>^2 & 0.5\times<\delta x_{\omega s}\delta y_{\omega s}+\delta y_{\omega s}\delta x_{\omega s}>-<\delta x_{\omega s}><\delta y_{\omega s}> \\ 0.5\times<\delta x_{\omega s}\delta y_{\omega s}+\delta y_{\omega s}\delta x_{\omega s}>-<\delta x_{\omega s}><\delta y_{\omega s}> & <\delta y_{\omega s}^2>-<\delta y_{\omega s}>^2 \end{bmatrix}$$

$$B = \begin{bmatrix} <\delta x_{\omega i}^2>-<\delta x_{\omega i}>^2 & 0.5\times<\delta x_{\omega i}\delta y_{\omega i}+\delta y_{\omega i}\delta x_{\omega i}>-<\delta x_{\omega i}><\delta y_{\omega i}> \\ 0.5\times<\delta y_{\omega i}\delta x_{\omega i}+\delta x_{\omega i}\delta y_{\omega i}>-<\delta y_{\omega i}><\delta x_{\omega i}> & <\delta y_{\omega i}^2>-<\delta y_{\omega i}>^2 \end{bmatrix} \quad (12)$$

$$C = \begin{bmatrix} 0.5\times<\delta x_{\omega s}\delta x_{\omega i}+\delta x_{\omega i}\delta x_{\omega s}>-<\delta x_{\omega s}><\delta x_{\omega i}> & 0.5\times<\delta x_{\omega s}\delta y_{\omega i}+\delta y_{\omega i}\delta x_{\omega s}>-<\delta x_{\omega s}><\delta y_{\omega i}> \\ 0.5\times<\delta y_{\omega s}\delta x_{\omega i}+\delta x_{\omega i}\delta y_{\omega s}>-<\delta y_{\omega s}><\delta x_{\omega i}> & 0.5\times<\delta y_{\omega s}\delta y_{\omega i}+\delta y_{\omega i}\delta y_{\omega s}>-<\delta y_{\omega s}><\delta y_{\omega i}> \end{bmatrix}$$

Therefore, using Eq. 12, and substituting into Eq. 9, one can study the entanglement between cavities modes. In the following, the results of the simulations are presented, and it will be discussed that how the designed sensor can duplicate the Robins.

**Results and Discussions**

This section presents the simulation results. Table 1 tabulates all the constant data that need to simulate the quantum sensor. In the following simulations, we use the external magnetic field coefficients indicated with $V_{H1}$ and $V_{H2}$. These coefficients, normalization factors, determine how much of the magnetic field can be sensed by Hall sensors. For instance, $V_{H1} = 1$ and $V_{H2} = 0$ mean that S-pole of the magnet field is completely faced toward the Hall sensor of S_I, and N-pole is faced toward S_II. Furthermore, $V_{H1} = 0$ and $V_{H2} = 0$ mean that there are no any magnetic fields affecting Hall sensors. Initially, the external magnetic field effect is considered on the entanglement between $OC_1$ and $OC_2$ modes. Fig. 1a shows that when the applied external field coefficients are equal, there is no any entangling between $OC_1$ and $OC_2$ modes. In fact, these results indicate that two initially entangled optical modes (signal and idler) become separable when they are interacted with the system, and the external field is employed. However, Fig. 1b shows that $OC_1$ and

$OC_2$ modes are entangled. In fact, it will be possible to have entanglement between the interacted optical cavities modes for some external field coefficients.

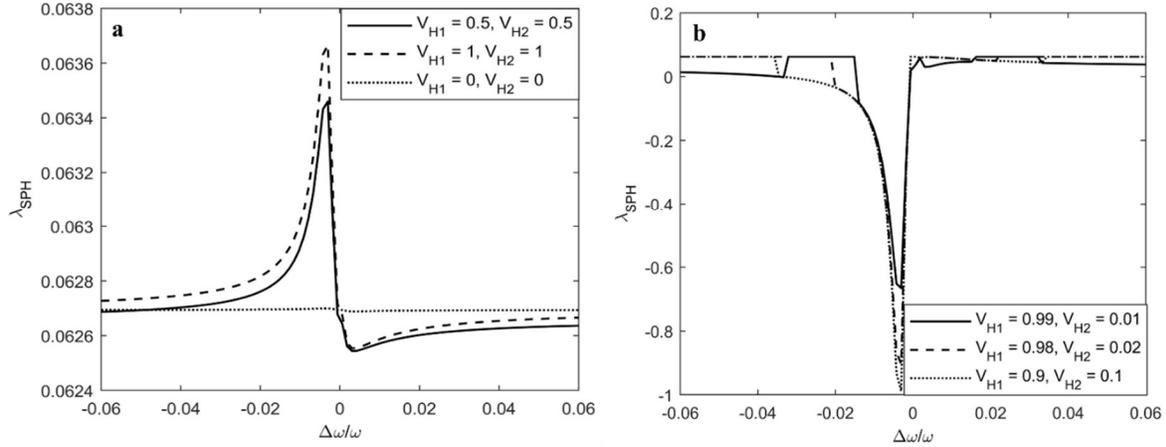

Fig. 1 External magnetic field coefficient effect on $OC_1$ and $OC_2$ modes entanglement; T = 350 mK, $\chi^{(2)}$ = $1.2*10^{-12}$ $m^2/v^2$.

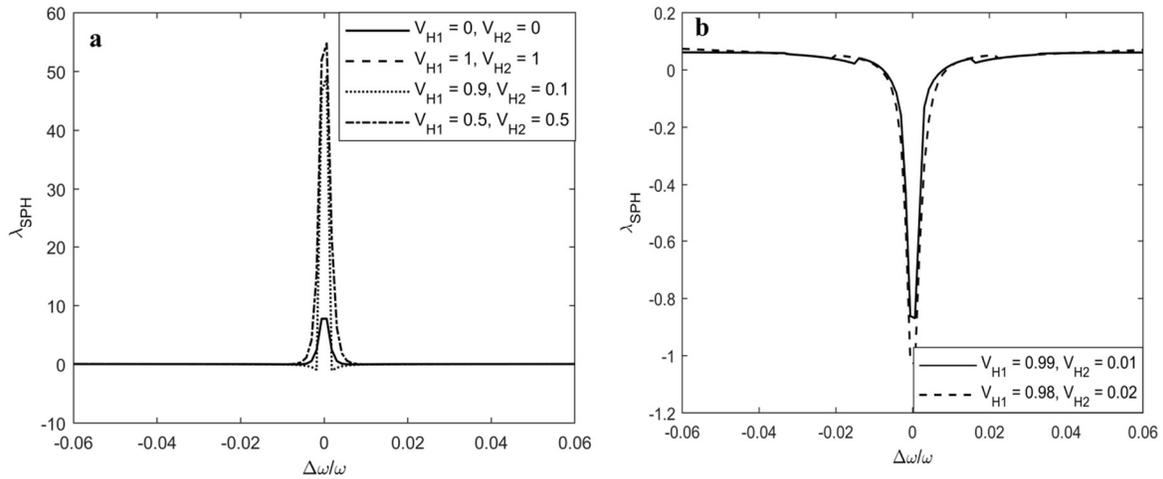

Fig. 2 External magnetic field coefficient effect on $MC_1$ and $MC_2$ modes entanglement; T = 350 mK, $\chi^{(2)}$ = $1.2*10^{-12}$ $m^2/v^2$

It is worthy to note that the quantum sensor is designed to detect the external magnetic field regarding the entanglement between $MC_1$ and $MC_2$ not $OC_1$ and $OC_2$; since the simulation results show that the entanglement between $MC_1$ and $MC_2$ is highly sensitive than $OC_1$ and $OC_2$ as the external magnetic field is changed. Therefore, we just focus on the entanglement between $MC_1$ and $MC_2$ modes. Fig. 2a shows that when the applied external field coefficients are equal, like ($V_{H1}$ = 1, $V_{H2}$ = 1), ($V_{H1}$ = 0, $V_{H2}$ = 0) or ($V_{H1}$ = 0.5, $V_{H2}$ = 0.5), the $MC_1$ and $MC_2$ modes are separable. Certainly, the two latter mentioned coefficients are not practical, since it is impossible to have two S-poles ($V_{H1}$ = 1, $V_{H2}$ = 1) or N-poles ($V_{H1}$ = 0, $V_{H2}$ = 0) on one magnetic field line. It is only an attempt to show that the prepared codes correctly work to simulate the entanglement between the modes. The interesting case used in this study is $V_{H1}$ = 0.99, $V_{H2}$ = 0.01, which means that 99 percent of S-pole is faced toward S_I, and the same percent of N-pole is faced toward S_II. In other words, 99 percent of the external magnetic field (S-pole) is applied on the Hall sensor of S_I. For the mentioned coefficients, Fig. 2b shows that the $MC_1$ and $MC_2$ modes are entangled. Moreover, the same thing occurs for $V_{H1} = 0.98$, $V_{H2} = 0.02$, whereas by decreasing $V_{H1}$ to 0.9, the two microwave cavities modes become separable as illustrated in Fig. 2a. In other words, the designed quantum sensor is highly sensitive to change of the angle of the applied magnetic field. Therefore, the small deviation from the line of the magnetic field that the sensor is put on, will be detected. The entanglement between $MC_1$ and $MC_2$ modes is an effective criterion utilized by the designed quantum sensor to detect the maximum intensity of applied field ($\alpha_e$ in scheme 2b). The entanglement between microwave cavities modes ($MC_1$ and $MC_2$) in the designed system seems to be similar with the process used by the Robins. In fact, it is the external magnetic field intensity disturbing the correlation between the modes as same as disturbing the dancing between singlet and triplet sates in the Robins. Considering Scheme 2b, if one accurately moves with a quantum senor on a magnetic field line, without changing the line, the two microwave cavities modes remained entangled. By changing the magnetic field line, if the voltage coefficients are decreased, i.e. $V_{H1} < 0.98$, then the two modes become separable. In other words, any changes of the direction of the magnetic field line lead to change the modes correlation behavior. Therefore, any deviation from the magnetic line can be detected by the designed quantum sensor.

After defining the criterion that the quantum sensor works with, now, it is important to study any parameters that can disturb the mentioned criterion. The important disadvantages of the entanglement are its unstability and fragility [2]. These features are easily accounted by the interaction of the quantum sensor with its operational environment. The highly critical factor is the noise effect addressed with n(s) in Eq. 7. From the correlation relation of the noise in Eq. 8, it is shown that this quantity is strongly affected by the environment temperature. Therefore, we study the effect of the temperature changing on the quantum sensor criterion, and Fig. 3 illustrates the results of the simulations. This figure shows that by increasing the temperature on order of 10 K, the $MC_1$ and $MC_2$ modes become completely separable (Fig. 3d). However, at the temperature below 1550 mK, it seems that the quantum sensor can operate safely when $\Delta\omega/\omega \sim 0$, otherwise, the operational temperature should be limited below 400 mK. The important reason leading to confine the operation of the quantum sensor at a very low temperature is the MR cavity operational frequency. In fact, it is due to the low frequency that MR oscillator works with ($\omega_m = 2\pi*10^6$ Hz); low frequency means high noise, and this point is shown by Eq. 8. To solve the problem, there are some studies utilizing different techniques either with engineering the frequency bandwidth [10, 11] or replacing the MR system with the optoelectronic device operating with a high frequency [12, 13]. The other important parameter that can affect the entanglement between two microwave modes is the second order susceptibility $\chi^{(2)}$ of the nonlinear OPDC used in the quantum sensor depicted in Fig. 4. It is shown that by decreasing $\chi^{(2)}$, the entanglement between microwave cavities modes is disturbed. The nonlinearity of the OPDC is decreased by decreasing $\chi^{(2)}$, which means that the entanglement property of the generated photons (signal and idler) before the interaction is decreased. Therefore, alteration of the signal and idler quantum property exciting the tripartite systems, affects the microwave cavities entanglement. Here, it can be concluded that the necessary condition to have entanglement between $MC_1$ and $MC_2$ is to initially establish the entangled signal and idler which is dependent on the OPDC nonlinearity factor $\chi^{(2)}$.

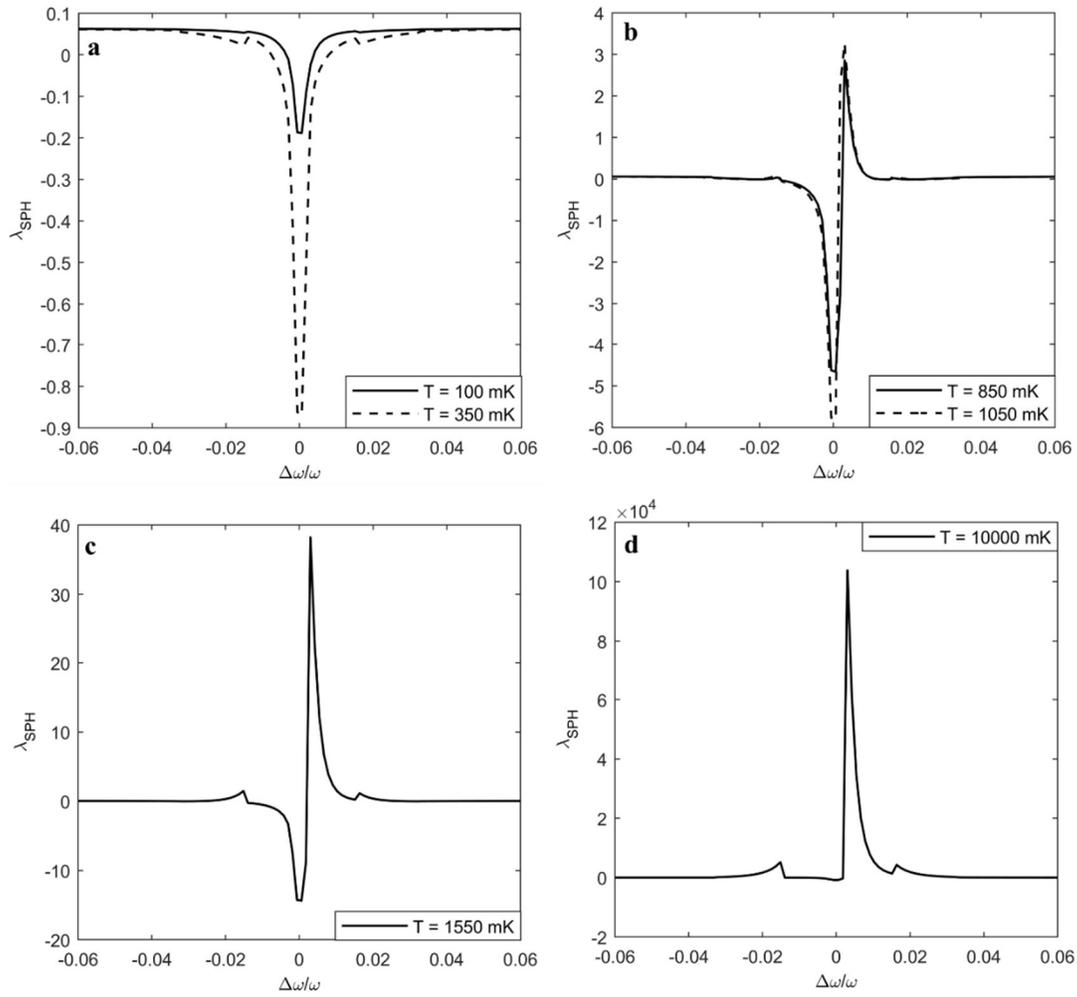

Fig. 3 Temperature effect on entanglement between $MC_1$ and $MC_2$ modes; $V_{H1} = 0.99$, $V_{H2} = 0.01$, $\chi^{(2)} = 1.2*10^{-12}$ $m^2/v^2$

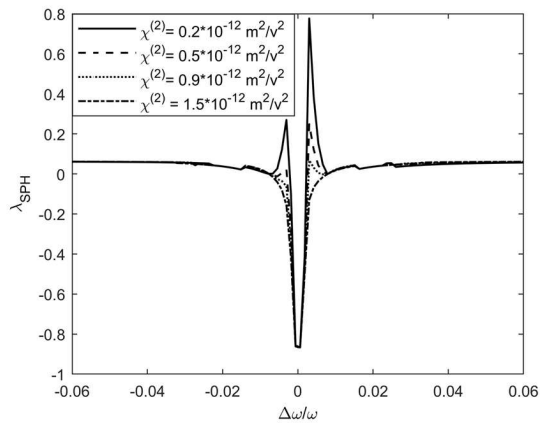

Fig. 4 Second order susceptibility effect on $MC_1$ and $MC_2$ modes entanglement; $V_{H1} = 0.99$, $V_{H2} = 0.01$, $T = 350$ mK

In the following, one of important feature of the quantum sensor is studied. In contrast to the Robins, the designed sensor is not affected by the other RF sources in the frequency range of <= 1.6 GHz. In [1], the results shown that at some critical frequencies on order of MHz, the operation of the Robins was disturbed. To make some comparisons to the Robins procedure, an external RF source is employed with a modified voltage coefficient as $V_{H1m} = V_{H1} + V_{H1}*\cos(2\pi f_{ext}t)*\cos(\theta_{ext})$ where $f_{ext}$ is the RF source frequency, and $\theta_{ext}$ is the angle that the RF source is excited respecting the original magnetic field. Fig. 5 shows the simulation results. First, the effect of the RF signal frequency on the Varactor diode capacitor is considered. As Fig. 5a shows, the Varactor diode capacitor reveals different behaviors due to emergence of some other elements, including R, $R_E$, and $L_E$ at RF-MW frequency range [26]. The alteration of the Varactor diode capacitance is studied, and Fig. 5b depicts the simulation results. It is shown that at the range of frequency <1 GHz, the capacitance is approximately constant and equals with $C_{VS}$. However, by increasing the frequency up to 1GHz, the Varactor diode capacitance is strongly affected by RF-MW frequencies. First, the modified capacitance is increased as the frequency approaches to resonance. Second, as the frequency continues to increase, the influence of the parasitic inductance becomes dominant and the modified capacitance is dramatically decreased. The change of the Varactor diode capacitance dramatically affects the LC circuit frequency. As Fig. 5c shows, the LC frequency shows a unique behavior at higher frequencies ($f_{ext}$> 1.6 GHz). Finally, as Fig. 5d shows, the effect of the RF source frequency on the microwave cavities modes correlation is investigated. This figure shows that by increasing the RF source frequency up to 1.6 GHz, the $MC_1$ and $MC_2$ modes become strongly separable. It contributes to the Varactor diode changing, and also the distortion of LC circuit frequency within the mentioned frequencies range.

**Conclusions**
In this article, a quantum sensor was designed to duplicate the Robins procedure. The quantum sensor had two distinct tripartite systems, which were separately excited by entangled photons. It was shown that the microwave cavities of the tripartite systems could couple nonclassicaly to each other. In fact, it was the main criterion in which the quantum sensor was employed to sense the small alteration of the applied magnetic field. It was a typical Hall sensor to sense the magnetic field intensity, and then the output signal was the voltage that drop across the Varactor diode. Then, the Varactor diode capacitance was changed based on the voltage drop. The change of the Varactor diode capacitance altered the LC circuit total capacitance. In other words, the LC circuit frequency was dependent on the alteration of the external magnetic field. The simulation results showed that the $MC_1$ and $MC_2$ modes could be entangled, and changing some parameters such as the magnetic field intensity, temperature and quantum property of the entangled photons strongly affected the quantum sensor operation. Consequently, the entanglement between $MC_1$ and $MC_2$ modes was the criterion utilized by the quantum sensor to duplicate the Robins procedure. It was shown that the entanglement between two microwave modes is severely influenced as same as the alteration of the singlet-triplet state dancing due to the geomagnetic field effect. Moreover, it was examined that the designed sensor was not affected by the external RF source at the range of MHz, which was in contrast to the Robins that was easily affected within the mentioned frequency range. It was notable to indicate that the quantum sensor was analyzed with the quantum electrodynamics theory in which the system dynamics equations of motions were derived using Heisenberg-Langevin equations.

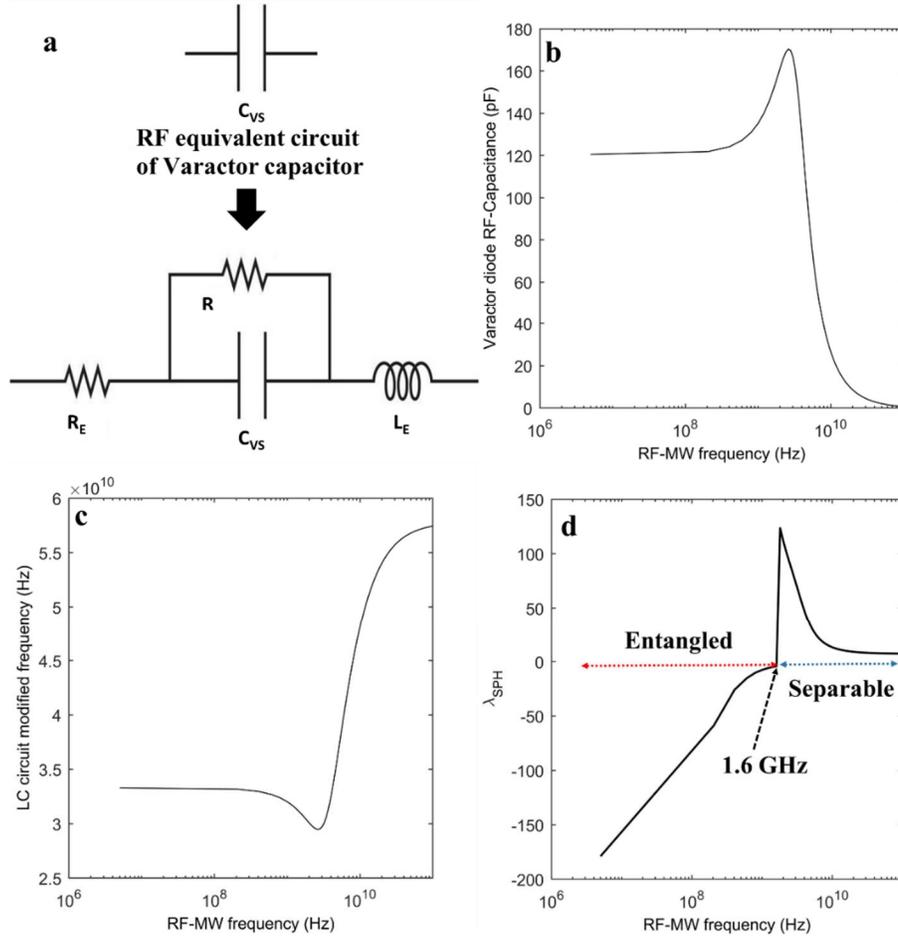

Fig. 5 Radio-Microwave (RF-MW) frequency effect on entanglement between $MC_1$ and $MC_2$ modes; $V_{H1}$ = 0.99, $V_{H1m} = V_{H1} + V_{H1}*\cos(2\pi f_{ext}t)*\cos(\theta_{ext})$, $V_{H2m} = 1 - V_{H1m}$, $\theta_{ext}$ = 73 deg, $C_{VS}$ = 120 pF, T = 350 mK, $\Delta\omega/\omega$ = 0.